\documentclass[12pt]{article}

\usepackage{graphicx}

\begin{document}

\begin{center}
{\LARGE IACTalks: an on-line archive of astronomy-related seminars}
\end{center}

\begin{center}
\large
Johan H. Knapen$^{1,2}$, Jorge A. P\'{e}rez Prieto$^{1}$, Tariq Shahbaz$^{1,2}$,
Anna Ferr\'{e}-Mateu$^{1,2}$, Nicola Caon$^{1,2}$, Cristina Ramos Almeida$^{1,2}$, 
Brandon Tingley$^{1,2}$, Valentina Luridiana$^{1,2}$, In\'{e}s Flores-Cacho$^{1,2,3,4}$,  
Orlagh Creevey$^{1,2,5}$,  Arturo Manchado Torres$^{1,2,6}$, Ignacio Trujillo$^{1,2}$, 
Maria Rosa Zapatero Osorio$^{1,2,7}$, Francisco S\'{a}nchez Mart\'{i}nez$^{1,2}$, 
Francisco L\'{o}pez Molina$^{1}$, Gabriel P\'{e}rez D\'{i}az$^{1}$, Miguel Briganti$^{1}$, 
In\'{e}s Bonet$^{1}$ 
\end{center}

\begin{center}

\noindent
\small $^{1}$Instituto de Astrof{\'\i}sica de Canarias, E-38205 La Laguna, Tenerife, Spain \\
\small $^{2}$Departamento de Astrof{\'\i}sica, Universidad de La Laguna, 
E-38206 La Laguna, Tenerife, Spain\\
\small $^{3}$CNRS, IRAP, 9 Av. colonel Roche, BP 44346, F-31028 Toulouse cedex 4, France \\
\small $^{4}$Universit{\'e} de Toulouse, UPS-OMP, IRAP, F-31028 Toulouse cedex 4, France \\
\small $^{5}$Laboratoire Lagrange, UMR 7293, CNRS (OCA), Universit\'e de Nice Sophia-Antipolis, Nice, France \\
\small $^{6}$Consejo Superior de Investigaciones Cient{\'i}ficas, Spain \\
\small $^{7}$Centro de Astrobiolog{\'i}a (CSIC-INTA), Ctra. Ajalvir km. 4, 28850 Torrej{\'o}n de Ardoz, Madrid, Spain
\end{center}

\date{\today}

\begin{abstract}
We present IACTalks, a free and open access seminars archive (http://iactalks.iac.es) aimed at promoting
astronomy and the exchange of ideas by providing high-quality scientific
seminars to the astronomical community. The archive of seminars and talks given
at the Instituto de Astrof{\'\i}sica de Canarias goes back to 2008. Over 360 talks
and seminars are now freely available by streaming over the internet. We
describe the user interface, which includes two video streams, one showing the
speaker, the other the presentation. A search function is available, and
seminars are indexed by keywords and in some cases by series, such as special
training courses or the 2011 Winter School of Astrophysics, on secular evolution of
galaxies. The archive is made available as an open resource, to be used by
scientists and the public.
\end{abstract}

\section{Introduction}

Since April 2008, the Instituto de Astrof{\'\i}sica de Canarias (IAC) has been
webcasting and archiving the weekly seminars and monthly colloquia about the
most exciting topics in astronomy and astrophysics' research.  The IAC is one of
the leading centres for astrophysical research and related technology
development in Spain, and has a staff of over 350, of which over 160 are active
in astronomy research.

The seminars cover both observational and theoretical astronomy and include
diverse topics such as particle and stellar physics, galaxies and cosmology, and
astronomical instrumentation and telescopes. The seminars are aimed at a general
astronomical audience and mainly given by postdocs and staff
astronomers. The speakers in our special ``colloquia'' are high-profile scientists
and over the years have included leading astronomers from all over the world,
covering practically the complete spectrum of current-day astrophysical
research.

While most of the archived seminars are on astrophysical research, the IACTalks
archive also contains seminars on various topics relating to astronomical
telescopes and instrumentation, and programming and (super)computing.

Other on-line seminars include the webcasting and archive service of the Space
Telescope Science Institute (https://webcast.stsci.edu/webcast/), going back to
2001, or the recently started seminar series at
http://asterisk.apod.com/ampersand/ .  IACTalks is complementary to these and
other initiatives.

\section{The IACTalks web archive}

\begin{figure} 
\centering
\includegraphics[scale=0.5]{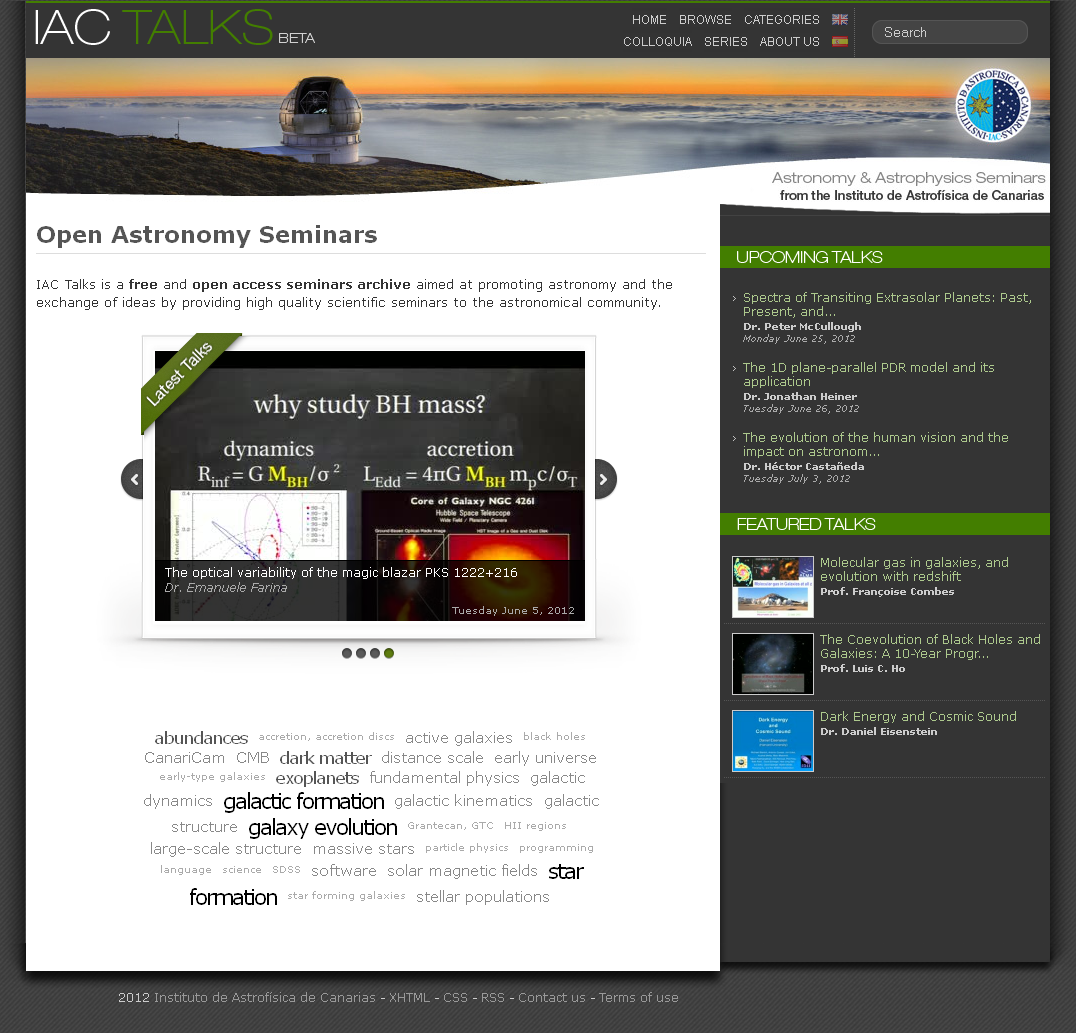}
\caption{Screenshot of the homepage}
\label{homepage}
\end{figure}

The archive, accessed via the web address http://iactalks.iac.es (see Fig. \ref{homepage}),
currently contains over 360 seminars, recorded  since April 2008. Of these, 45
form part of the ``colloquium'' series, for which the IAC invites renowned
speakers from anywhere in the world to spend a week at the institute to interact
with staff and students, and to give a seminar.

The various seminars are organized in 13 categories, and each seminar is
labelled with a number of keywords. The web interface allows browsing through
the available talks, searching for names or keywords, or browsing by categories
(such as galaxies, planetary systems, or the Sun) or series. The latter include
recordings of workshops and courses, including the XXIIIrd Canary Islands Winter
School, held in November 2011 in Tenerife on the topic of secular evolution in
galaxies.

\begin{figure}
\centering
\includegraphics[scale=0.5]{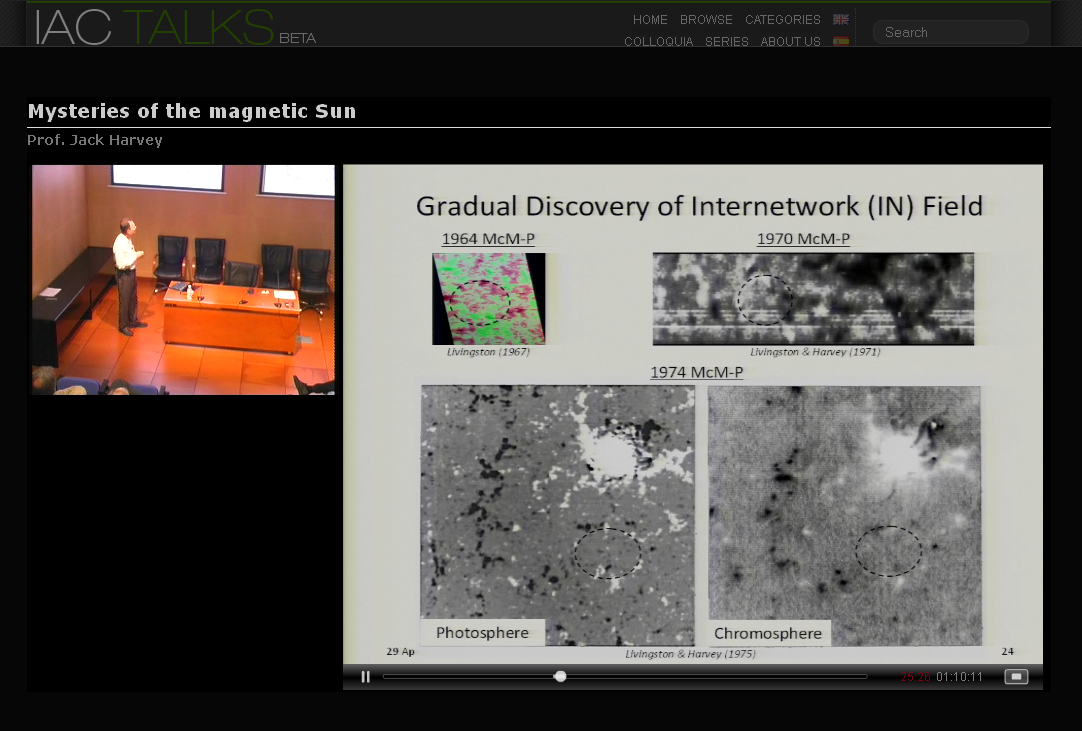}
\caption{Screenshot of the appearance of a talk during playback.}
\label{talk}
\end{figure}

Seminars at the IAC are broadcast over the internet and can be watched either
live or in the archived form described here. We capture a direct video feed of
the slides as they appear on the screen and a fixed camera also films the
speaker during the presentation. Both these videos appear on the webcast, and
either can be enlarged to fill the screen of the viewer. Figure \ref{talk} shows a
screenshot of the appearance during playback.  We use a web-based player
(Flowplayer), so only a HTML5 compliant browser with Adobe Flash is required.

\section{Future improvements}

While our current system should work without problems on most browsers, in the
near future we plan to release a specific interface for full compatibility with
mobile devices, including Apple hand-held devices which do not support Flash
technology. We also plan to release integration to social networks.

In addition, we are considering using free on-line services (such as those
provided by Google and YouTube) to provide automatically-generated transcripts
of all seminars, which can then also be translated automatically into a multitude
of languages, and/or superimposed as subtitles. While this will not guarantee
accurate reflections of everything a speaker says, especially when jargon is
involved, our initial tests have shown that the results of this exercise can
make the seminars available to a huge world-wide audience of interested amateur
astronomers who lack the language skills to follow a seminar in English, but can
get the essence of a presentation when it is supported by subtitles in their
own language.

\section{Conclusions}

This short note is aimed to announce the launch of the IACTalks website,
http://iactalks.iac.es, which yields free and open access to a large archive of
seminars, aimed at promoting astronomy and the exchange of ideas by providing
high-quality scientific seminars to the astronomical community. The archive of
seminars and talks given at the Instituto de Astrof{\'\i}sica de Canarias goes back
to 2008, and now contains over 360 talks and seminars available by streaming
over the Internet. We have described the main features of the website, and
outlined future plans for further development.

\end{document}